\documentclass[12pt]{iopart} 
\usepackage{epsf}
\usepackage{graphicx} 
\begin{document}
\letter{Soft versus Hard Dynamics for Field-driven Solid-on-Solid Interfaces}
\author{
P A Rikvold$\dag\ddag$ and M Kolesik$\S\|$
}
\address{
$\dag$ Center for Materials Research and Technology, 
School of Computational Science and Information Technology, and 
Department of Physics,
Florida State University, Tallahassee, Florida 32306-4351, USA\\ 
$\ddag$ Department of Fundamental Sciences, Faculty of Integrated Human Studies, 
Kyoto University, Kyoto 606, Japan\\
$\S$Institute of Physics, Slovak Academy of Sciences,
Bratislava, Slovak Republic\\ 
$\|$ Department of Mathematics, University of Arizona,
Tucson, Arizona 85721, USA
}

\eads{\mailto{rikvold@csit.fsu.edu}, \mailto{kolesik@acms.arizona.edu}}

\begin{abstract}
Analytical arguments and dynamic Monte Carlo simulations 
show that the microscopic structure of field-driven Solid-on-Solid interfaces 
depends strongly on the details of the dynamics. For nonconservative
dynamics with transition rates that factorize into parts dependent 
only on the changes in interaction energy and field energy, respectively 
(soft dynamics), the intrinsic interface width is field-independent. 
For non-factorizing rates, 
such as the standard Glauber and Metropolis algorithms (hard dynamics), it 
increases with the field. 
Consequences for the interface velocity and its anisotropy are discussed.
\end{abstract}

\submitto{\JPA}

\maketitle

\nosections
Surfaces and interfaces moving under far-from-equilibrium conditions 
are important in 
the formation of patterns and structures. For example, domain growth by the 
motion of defects such as grain boundaries or dislocations 
influences the mechanical properties of metals \cite{MEND00} and 
the structures formed during phase transformations in 
adsorbate systems \cite{MITC01}, 
while the propagation of domain walls influences the switching dynamics of 
magnetic nanoparticles and ultrathin films \cite{RIKV02A}. 
Recently, interface dynamics was even used to analyze the 
scalability of discrete-event simulations on parallel computers \cite{KORN00}. 
The importance of moving interfaces in a wide variety of 
fields has inspired enormous interest 
in their structure and dynamics \cite{BARA95}. 
However, despite the fact that 
many interface properties, such as mobility and chemical activity, 
are determined by the {\it microscopic\/} structure, 
interest has mostly focused on the large-scale interface structure 
and its universal properties. In this Letter we show that 
the microstructure of a moving interface 
can be dramatically influenced by seemingly minor modifications of 
the growth mechanism, 
with measurable consequences for such macroscopic quantities as 
the interface velocity and its anisotropy. 

The detailed microscopic mechanism of the interface motion is usually 
not known, and it is therefore common to mimic the essential 
features in a dynamic Monte Carlo (MC) simulation of a model stochastic 
process \cite{MEND00}. In doing so, there are two important distinctions that 
must be made. One is between those 
transition probabilities that only depend on the 
energy difference between the initial and final states (often referred to as 
Metropolis, Glauber, or heat-bath rates \cite{SIEG94,LAND00,SHIM01}), 
and those that involve an activation barrier between the two states 
(often referred to as Arrhenius rates) \cite{MITC01,SIEG94,SHIM01,KANG89}. 
Another distinction is between dynamics that do not 
conserve the order parameter, such as the Metropolis and Glauber 
single-spin flip algorithms, and conservative dynamics, 
such as the Kawasaki spin-exchange dynamic \cite{LAND00}. 
Once it is decided to which of these categories the 
system belongs, the dynamic is often chosen on the basis of convenience.

It is well known that different microscopic dynamics can yield 
different equilibration paths and equilibrium fluctuations \cite{SHOC92}
(cluster vs local MC algorithms being the most extreme example
\cite{LAND00}) and even noticeable differences in the steady-state 
microstructure \cite{SIEG94,SHIM01}. 
Nevertheless, the general expectation is that, if no additional
parameters (such as an activation barrier or a diffusion rate) is
introduced into the physical model, observables  
are only affected {\it quantitatively\/}.
However, here we demonstrate striking {\em qualitative\/} differences 
between the microstructural consequences of two 
different stochastic dynamics that do not involve a transition barrier, even 
though both obey detailed balance and neither involves order-parameter
conserving moves. For this demonstration we use an unrestricted Solid-on-Solid 
(SOS) interface \cite{BURT51} 
separating uniform spin-up and spin-down phases in a 
ferromagnetic $S=1/2$ Ising system on a square lattice of unit lattice constant. 
The interface consists of integer-valued 
steps parallel to the $y$-direction, $\delta(x) \in \{ -\infty , +\infty \}$, 
where the step position 
$x$ is also an integer. Its energy is given by the 
nearest-neighbour Ising Hamiltonian with anisotropic ferromagnetic 
interactions, $J_x$ and $J_y$ in the $x$ and $y$ direction, respectively: 
${\cal H} = -\sum_{x,y} s_{x,y} \left( J_x s_{x+1,y} + J_y s_{x,y+1} 
+ H \right)$.
The two states at site $(x,y)$ are denoted by the spin $s_{x,y} = \pm 1$, 
$\sum_{x,y}$ runs over all sites, and $H$ is the applied field. We take 
$s_{x,y}=-1$ on the side of the interface corresponding to large positive $y$, 
so that the interface on average moves in the positive $y$ direction for 
$H > 0$. In the equivalent lattice-gas language, $s_{x,y} = +1$ 
corresponds to solid and $s_{x,y} =-1$ to gas or liquid, 
and $H$ is proportional to a chemical-potential imbalance. 
The interface evolves through single spin flips 
(in lattic-gas language: adsorption/desorption events) that occur 
with probability $W[\beta \Delta E]$ where $\Delta E$ is the energy change that 
would result from the transition, constrained by the requirement 
that the probabilities obey detailed balance 
(see \cite{RIKV00,RIKV02} for details). 
To restrict the accessible configurations to 
a simple SOS interface without bubbles or overhangs, 
transitions are allowed only for spins that have one single broken
bond in the $y$-direction. 
Regardless of the details of the transition probabilities, which lead to 
dramatic {\it microscopic\/} structural differences as
demonstrated below, the {\it macroscopic\/} structure of the 
interfaces belongs to the Kardar-Parisi-Zhang (KPZ) dynamic universality class 
\cite{BARA95,KARD86}. 
We emphasize that the absence of energy barriers and 
order-parameter conserving diffusion (spin-exchange)
moves sets our model clearly apart from 
commonly studied models of Molecular Beam Epitaxy (MBE) \cite{SIEG94,SHIM01}. 
It is closely similar to the diffusion-free 
dynamical model studied in three dimensions in 
Ref.~\cite{JIAN92} which, however, does not consider the microscopic 
interface structure in detail. 

Recently \cite{RIKV00} we introduced a mean-field approximation for the 
driven-interface microstructure, in which 
the probability density function (pdf) for the 
height of a single step takes the 
Burton-Cabrera-Frank (BCF) form \cite{BURT51}, 
\begin{equation}
p[\delta(x)] = Z^{-1} X^{|\delta(x)|}
\ e^{ \gamma(\phi) \delta(x) } \;, 
\label{eq:step_pdf}
\end{equation}
where the Lagrange multiplier $\gamma(\phi)$ enforces 
$\langle \delta(x) \rangle = \tan \phi$ independent of $x$, 
corresponding to an overall angle $\phi$ between the interface and the $x$-axis. 
The width parameter $X$, which for a driven 
interface can depend on both $H$ and the temperature $T$, is discussed below. 
The partition function in (\ref{eq:step_pdf}) is 
\begin{equation}
Z
=
\sum_{\delta = -\infty}^{+\infty} 
X^{|\delta|} e^{ \gamma(\phi) \delta } 
= 
\frac{1-X^2}{1 - 2 X \cosh \gamma(\phi) + X^2} 
\label{eq:Z}
\end{equation}
with
\begin{equation}
e^{\gamma (\phi)} 
= 
\frac{ \left(1+X^2 \right)\tan \phi 
+ 
\sqrt{ \left( 1 - X^2 \right)^2 \tan^2 \phi + 4 X^2 }
}
{2 X \left( 1 + \tan \phi \right)} 
\;.
\label{eq:chgam}
\end{equation}
For $\phi = 0$, $\gamma(\phi) = 0$, yielding $Z(0) = (1+X)/(1-X)$. 

In the approximation of \cite{RIKV00}, 
individual steps are assumed to be statistically independent (like 
in the original BCF model for {\it equilibrium\/} interfaces). 
The width parameter $X$ is then obtained self-consistently as 
\cite{RIKV00,RIKV02} 
\begin{equation}
\hspace{-2cm}
X(T,H) = 
X_0(T) 
\left\{
\frac{e^{-2 \beta H}W[\beta(-2H-4J_x)] + e^{2\beta H}W[\beta(+2H-4J_x)]}
{W[\beta(-2H-4J_x)] + W[\beta(+2H-4J_x)]}
\right\}^{1/2}
\;,
\label{eq:XTH}
\end{equation}
where $X_0(T) \equiv e^{-2 \beta J_x}$ with $\beta = 1/k_{\rm B}T$ ($k_{\rm B}$ 
is Boltzmann's constant) is the BCF zero-field equilibrium value of $X$. 
For $H=0$, $X(T,H)$ reduces to $X_0(T)$.
For instance, using the Glauber transition probability, 
\begin{equation}
W_{\rm G}(s_{x,y} \rightarrow -s_{x,y}) 
= \frac{e^{-\beta \Delta E }}{1 + e^{-\beta \Delta E }} 
\;,
\label{eq:glau}
\end{equation}
we get 
\begin{equation}
X_{\rm G}(T,H) = 
X_0(T)
\left\{
\frac{e^{2 \beta J_x} \cosh(2 \beta H) + e^{-2\beta J_x}}
{e^{-2 \beta J_x} \cosh(2 \beta H) + e^{2\beta J_x}}
\right\}^{1/2}
.
\label{eq:XG}
\end{equation}

Equation (\ref{eq:glau}) exemplifies a class of dynamics known as 
{\it hard\/} in the literature on nonequilibrium lattice models \cite{MARR99}, 
in which the transition probabilities depend directly on the 
total energy change, $\Delta E$. In a different class, known as 
{\it soft\/}, the probabilities factorize into a part due 
only to the change in the field energy, $\Delta E_H \propto H$, and a part due 
only to the change in the interaction energy, $\Delta E_J \propto J_x$. 
One example is the {\it soft Glauber dynamic\/}, 
\begin{equation}
W_{\rm SG}(s_{x,y} \rightarrow -s_{x,y}) 
= 
\frac{e^{-\beta \Delta E_H }}{1 + e^{-\beta \Delta E_H }} 
\cdot
\frac{e^{-\beta \Delta E_J }}{1 + e^{-\beta \Delta E_J }} 
\;,
\label{eq:SG}
\end{equation}
which also obeys detailed balance.

The important point is that when a factorizing 
transition probability such as (\ref{eq:SG}) 
is used in (\ref{eq:XTH}), all dependence on $H$ cancels 
due to the detailed balance. 
Whether one uses a soft or a hard dynamic thus strongly affects 
the intrinsic interface width and properties that depend 
on it, such as the propagation velocity. 
The need to use a soft dynamic in cases where the ``field'' represents 
a chemical-potential difference has been recognized in some MC studies of 
crystal growth \cite{GROS91,KOTR91,HONT97}. 

Next we illustrate the large differences between hard and soft 
dynamics by comparing analytic approximations and dynamic 
MC simulations of driven SOS interfaces with $J_x$=$J_y$=$J$, using the hard and 
soft Glauber dynamics, respectively. In particular, we consider the field- 
and temperature dependencies of the intrinsic interface width, 
represented by the average absolute value of the step height, 
and the resulting interface velocity and its anisotropy.

\begin{center}
\begin{figure}
\includegraphics[width=.50\textwidth,angle=270]{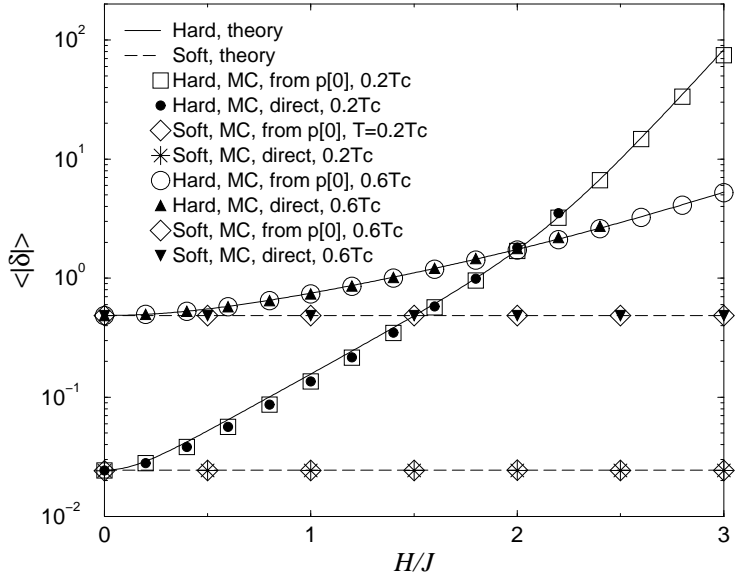}
\caption[]{
Average step height 
$\langle | \delta | \rangle$ vs $H$ for $\phi$=0 
at $T$=$0.2T_c$ and~0.6$T_c$, where $T_c$ is the exact 
Ising critical temperature. 
Solid curves, obtained as 
$\langle | \delta | \rangle = 2X/(1-X^2)$
with $X=X_{\rm G}(T,H)$, are theoretical predictions for the hard dynamic. 
Dashed lines, similarly obtained with 
$X=X_0(T)$, are theoretical predictions for the soft dynamic.
The MC data are from direct summation 
over the simulated single-step pdfs (filled symbols) and from $p[0]$ using 
$\langle | \delta | \rangle = \left\{ p[0]^{-1} - p[0] \right\}/2$ 
(empty symbols). 
In this and the following figures, the statistical uncertainty is 
much smaller than the symbol size. 
}                                  
\label{fig:meandelta}
\end{figure}
\end{center}

In figure~\ref{fig:meandelta} we show 
numerical evidence supporting our predictions for the average step height, 
which for $\phi = 0$ is related to $X$ as 
$\langle | \delta | \rangle = 2X/(1-X^2)$, 
under both the hard and soft Glauber dynamics at two different temperatures. 
(Simulational details are given below.) 
The difference between the two dynamics is 
striking: the step heights for the soft dynamic are 
independent of $H$, in contrast to the strong $H$-dependence 
produced by the hard dynamic. Additional confirmation of the functional form of 
the single-step pdf, (\ref{eq:step_pdf}), is obtained from the MC 
data by calculating  
$\langle | \delta | \rangle$, both directly by summation over the 
numerically obtained pdf, and from the probability 
of zero step height for $\phi = 0$ as 
$ \langle | \delta | \rangle = \left\{ p[0]^{-1} - p[0] \right\}/2$.
This result is obtained by combining $p[0] = Z^{-1}$ with the above relation 
for $\langle | \delta | \rangle$ in terms of $X$. 

The spins in an Ising system can be 
classified by the value of $\Delta E$, which is uniquely determined by $s$ and 
the number of broken $x$- and $y$-bonds. 
The active spin classes all have one broken $y$-bond and can 
be labeled by $s$=$\pm1$ 
and the number of broken $x$-bonds, $j \in \{0,1,2\}$, as $js$. 
The classes and corresponding $\Delta E$ are given in the 
first two columns of table~\ref{table}. 
The details of our ``active-site'' implementation 
of the spin-class based discrete-time $n$-fold way algorithm 
\cite{BORT75} 
are given in \cite{RIKV00}, the only difference here being the 
restriction to flips only at sites with one broken $y$-bond. 
The data shown here are for interfaces of size $L_x$=$10^4$ in the 
$x$-direction, which were allowed $5\,000$ $n$-fold way updates per active site
to reach stationarity, after which pdfs, class populations, and velocities were 
collected over $50\,000$ updates per active site. 

\begin{center}
\begin{figure}
\includegraphics[width=.50\textwidth,angle=270]{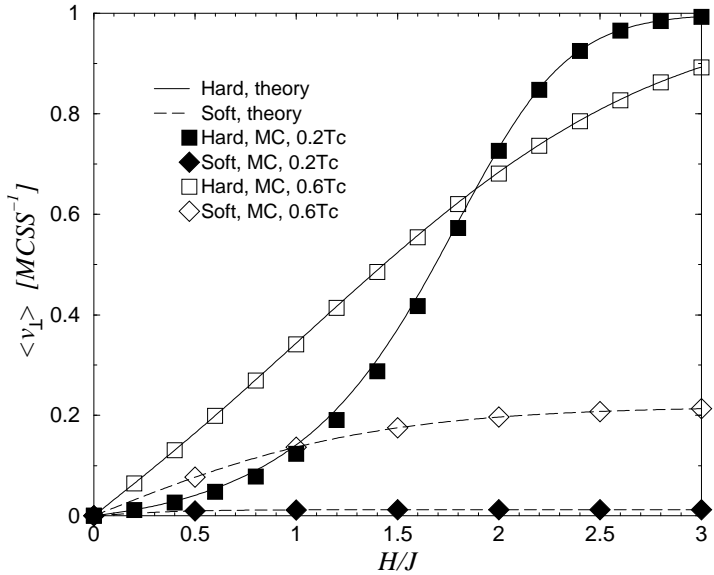}
\caption[]{
The perpendicular interface velocity $\langle v_\perp \rangle$ 
vs $H$ for $\phi = 0$ at $T$=$0.2T_c$ and~0.6$T_c$. Theoretical results 
for the hard dynamic are from (\ref{eq:totalv0G}) 
(solid curves), with MC data shown as squares. 
Theoretical results for the soft dynamic are from 
(\ref{eq:totalv0SG}) 
(dashed curves), with MC data shown as diamonds. 
}                                  
\label{fig:VvsH}
\end{figure}
\end{center}

Assuming up-down symmetry of the interface, 
the mean perpendicular propagation velocity is given by 
\begin{equation}
\langle v_\perp (T,H,\phi) \rangle 
= 
\cos \phi  \sum_{j} \langle n(js) \rangle \langle v_y(j) \rangle 
\label{eq:totalv}
\end{equation} 
in units of inverse MC steps per site (MCSS$^{-1}$). 
Here, the average spin-class populations, $\langle n(js) \rangle$, are 
calculated from the single-step pdf, assuming statistical independence 
\cite{RIKV00}, and 
the results are given in the third and fourth columns of table~\ref{table}. 
The corresponding contributions to the interface velocity in the $y$-direction, 
$\langle v_y(j) \rangle = \left\{ W \left[ \beta \Delta E(j-)  \right] 
- W \left[ \beta \Delta E(j+)  \right] \right\}$, are given in 
the fifth and sixth column of table~\ref{table} 
for the hard and soft Glauber dynamic, 
respectively. The special case of $\phi=0$ leads to  
compact formulas: 
\begin{equation}
\langle v_\perp (T,H,0) \rangle _{\rm G} 
=
\frac{\tanh (\beta H)}{(1+X)^2} 
\left\{
2 X 
+ 
\frac{1+X^2}
{1 + \left[\frac{\sinh (2 \beta J_x)}{\cosh (\beta H)}\right]^2} 
\right\} 
\label{eq:totalv0G}
\end{equation}
with $X$ from (\ref{eq:XG}) for the hard Glauber dynamic, and 
\begin{equation}
\langle v_\perp (T,H,0) \rangle _{\rm SG}
=
\tanh (\beta H) \frac{X_0}{1+X_0^2} 
\label{eq:totalv0SG}
\end{equation}
for the soft Glauber dynamic. 
These results are compared with the corresponding MC data in 
figure~\ref{fig:VvsH}. The difference between the 
two dynamics is clear. 

\begin{center}
\begin{figure}
\includegraphics[width=.50\textwidth,angle=270]{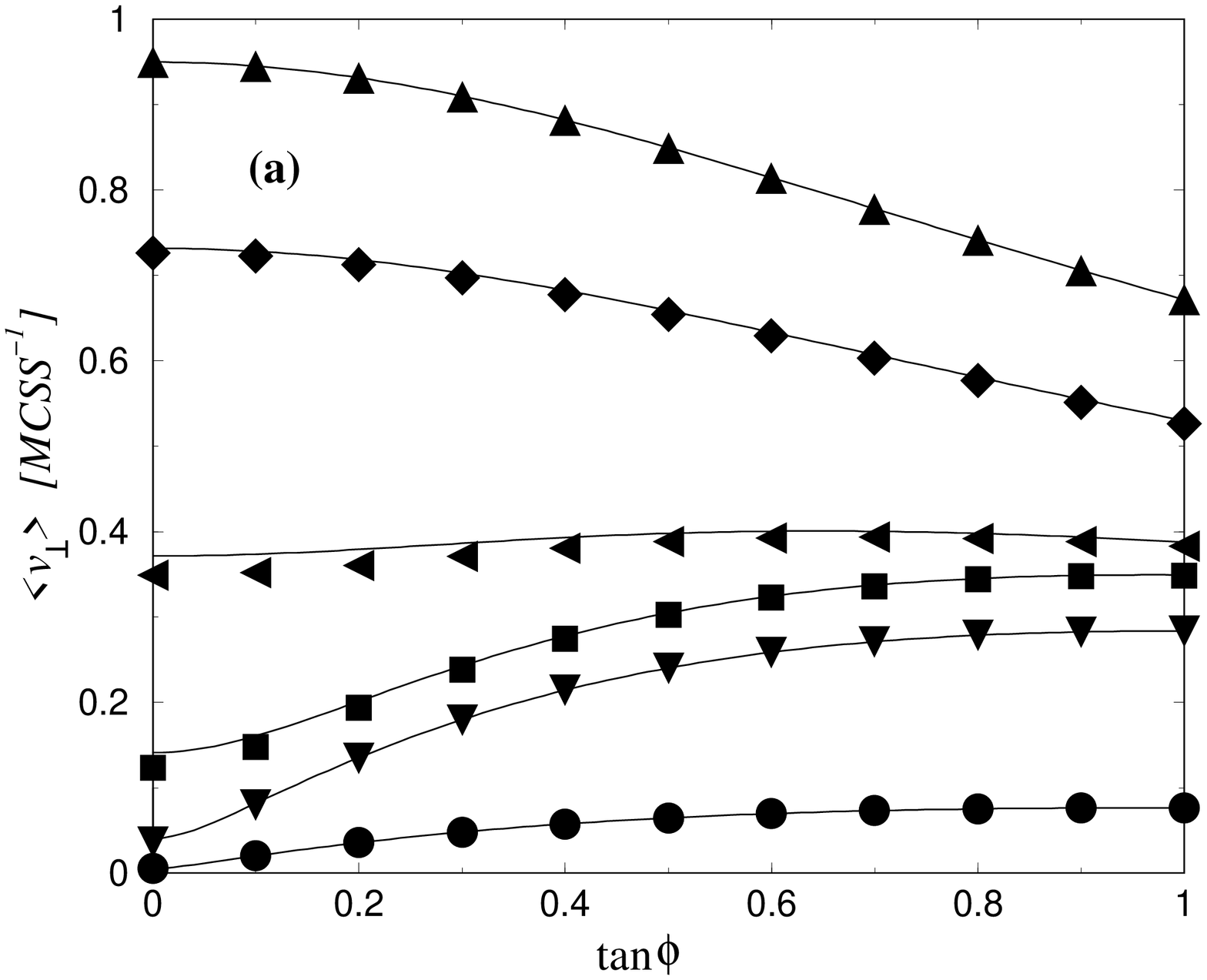}\\
\includegraphics[width=.50\textwidth,angle=270]{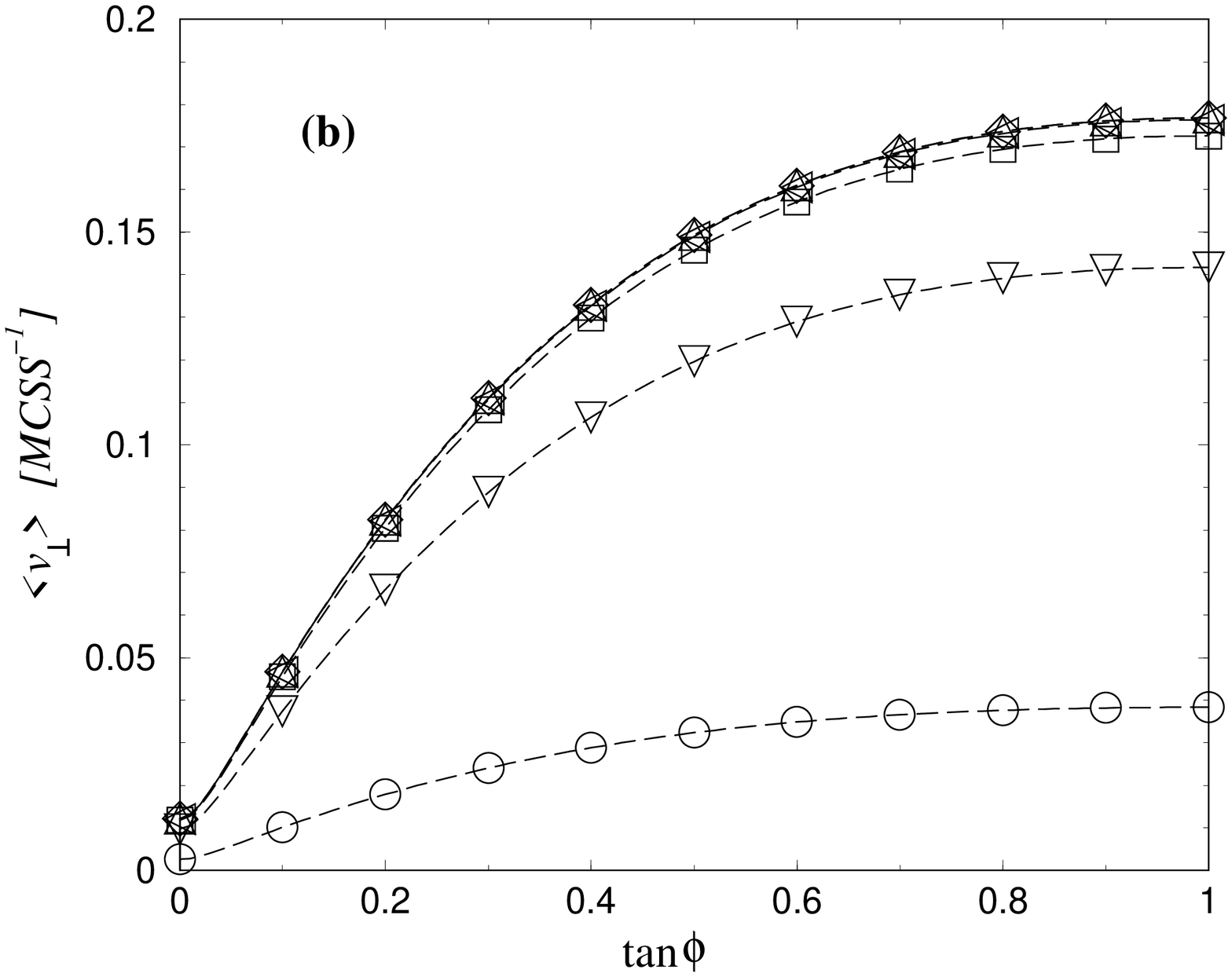}
\caption[]{
Angle dependence of the perpendicular 
interface velocity $\langle v_\perp \rangle$ 
at $T$=0.2$T_c$ for (from below to above) $H/J=0.1$, 0.5, 1.0, 1.5,
2.0, and 2.5. Curves are theoretical predictions, data points MC results. 
(a)
Hard dynamic.
(b)
Soft dynamic. On this scale, the soft-dynamic results overlap for $H/J \ge 1.5$. 
}                                  
\label{fig:VvsA}
\end{figure}
\end{center}

The angular dependencies of $\langle v_\perp \rangle$ 
for the two dynamics are illustrated in figure~\ref{fig:VvsA}. 
The theoretical results represent (\ref{eq:totalv}) with terms from 
table~\ref{table}. The hard dynamic yields an anisotropy which 
changes from that of the single-step model \cite{BARA95} for weak $H$, to the 
reverse anisotropy of the Eden model \cite{MEAK86B,HIRS86} 
for strong $H$. There is no such change for the soft dynamic, which shows a 
single-step like anisotropy for all fields. 
Effects of different transition probabilities on nonequilibrium
growth shapes have already been observed in systems with conserved order 
parameter \cite{SHOC92}. 
It is likely that the different anisotropies resulting from the hard and
soft dynamics will lead to different growth shapes in the case of
nonconserved order parameter, as well. 

With hard dynamics, the up-down symmetry of the interface is 
gradually destroyed 
with increasing field \cite{KORN00,RIKV00,RIKV02,NEER97}, 
leading to different class populations for $s$=+1 and 
$s$=$-$1. Our MC simulations show this skewness to be absent for the 
soft dynamic. 
This makes us believe that the 
field dependence in (\ref{eq:totalv0SG}) is exact in the following 
sense. If the interface has up-down symmetry
and is driven with a soft dynamic, then the generated interface shapes
are independent of $H$. Consequently, the only dependence 
of the velocity on $H$ is in the difference between the spin-flip probabilities 
of a down spin and an up spin, which yields
$\tanh(\beta H)$ for the Glauber dynamic.

In summary, we have presented analytical and dynamic MC results that show 
remarkable qualitative
differences in the microscopic structure of field-driven SOS 
interfaces in the KPZ dynamic universality class, 
depending on whether the local stochastic spin dynamic is hard (such as 
the standard Glauber and Metropolis dynamics) or soft. 
In the former case, $\langle | \delta | \rangle$ depends strongly on $H$,
while in the latter it has no $H$-dependence at all. 
We emphasize that the difference between the hard and soft dynamic is
purely dynamical and does {\it not\/} 
involve introduction into the physical model of any new parameters, 
such as an activation barrier or a diffusion rate. 
If microstructural information is desired from a dynamic MC simulation, 
great care must therefore be exercised, both in choosing the transition 
probabilities and in interpreting the results. 
Analogous effects for other models,  
such as Ising interfaces \cite{RIKV00}, and in three dimensions,
and the possibility that the resulting growth shapes may be different 
\cite{SHOC92}, are 
left for future study. Further details on the microstructure of the
SOS interface under the 
hard Glauber dynamic will be reported elsewhere \cite{RIKV02}.

\ack
We thank G~Brown, G~Buend{\'\i}a, S~J Mitchell, M~A Novotny, and K~Park 
for useful comments.
P~A~R appreciates hospitality at Kyoto University. 
Supported in part by US NSF Grant No.~DMR-9981815, and by Florida State 
University through MARTECH and CSIT.

%
%
\begin{table}[ht]
\caption[]{
The three spin classes that contribute to the interface velocity in 
the SOS model 
(column 1), together with 
the corresponding energy changes resulting from a successful 
spin flip (column 2, 
with upper sign corresponding to initial $s =-1$ and lower sign to $s=+1$), 
the average class populations for general tilt angle $\phi$ (column 3) 
and for $\phi = 0$ (column 4), 
and the contributions to the interface velocity in the $y$-direction 
for the hard Glauber dynamic (column 5) and for the soft Glauber dynamic 
(column 6). 
In columns 3 and 4, $X$ corresponds to the width parameter $X(T,H)$ for the 
specific dynamic used, given by (\ref{eq:XTH}) in the general case, 
(\ref{eq:XG}) for the hard Glauber dynamic, and 
$X_0(T) = e^{-2 \beta J_x}$ for soft dynamics. 
}
\label{table1}
\begin{tabular}{@{} l  l  l  l  l  l }
\br
Class 
& $\Delta E $
& $\langle n(js) \rangle$, 
& $\langle n(js) \rangle$, 
& $\langle v_y(js) \rangle_{\rm G}$ 
& $\langle v_y(js) \rangle_{\rm SG}$ 
\\ 
& 
& general $\phi$ 
& $\phi=0$ 
&  
& 
\\ 
 \mr
 $0s$ 
 & $\mp 2H +4J_x$
 & $\frac{1 - 2X \cosh \gamma (\phi) + X^2}{(1-X^2)^2}$ 
 & $\frac{1}{(1+X)^2}$ 
 & $\frac{\tanh\left( \beta H \right)}
	 {1 + \left[\frac{\sinh \left(2 \beta J_x \right) }
		   {\cosh \left( \beta H \right)}\right]^2 }$  
 & $\frac{\tanh\left( \beta H \right) X_0^2}{1 + X_0^2}$
\\
 $1s$ 
 & $\mp 2H$
 & $\frac{2X[(1+X^2) \cosh \gamma (\phi) - 2X]}{(1-X^2)^2}$ 
 & $\frac{2X}{(1+X)^2}$ 
 & ${\tanh\left( \beta H \right)}$
 & $\frac{1}{2} {\tanh\left( \beta H \right)}$
\\
 $2s$ 
 & $\mp 2H -4J_x$
 & $\frac{X^2[1-2X\cosh\gamma(\phi)+X^2]}{(1-X^2)^2}$ 
 & $\frac{X^2}{(1+X)^2}$ 
 & $\frac{\tanh\left( \beta H \right)}
	 {1 + \left[ \frac{\sinh \left(2 \beta J_x \right) }
		   {\cosh \left( \beta H \right)}\right]^2 }$  
 & $\frac{\tanh\left( \beta H \right)}{1 + X_0^2}$
\\
\br
 \end{tabular}
\label{table}
\end{table}

\clearpage


\section*{References}
\begin{thebibliography}{99}

\bibitem{MEND00}
Mendelev M~I and Srolovitz D~J 2000 {\it Acta Mater.\/} {\bf 48} 3711; 
2001 {\it ibid\/} {\bf 49} 589;
2001 {\it ibid\/} {\bf 49} 2843

\bibitem{MITC01}
Mitchell S~J, Brown G and Rikvold P~A 2001
{\it Surf.\ Sci.\/} {\bf 471} 125

\bibitem{RIKV02A}
Rikvold P~A, Brown G, Mitchell S~J and Novotny M~A 
in press {\it Nanostructured Megnetic Materials and their Applications\/} 
ed D~Shi, B~Aktas, L~Pust and F~Mikailov 
(Berlin: Springer) 
(preprint cond-mat/0110103) 

\bibitem{KORN00}
Korniss G, Toroczkai Z, Novotny M~A and Rikvold P~A
2000 {\it Phys.\ Rev.\ Lett.\/} {\bf 84} 1351  

\bibitem{BARA95}
Barab{\'a}si A-L and Stanley H~E
1995 {\em Fractal Concepts in Surface Growth\/}
(Cambridge: Cambridge Univ.\ Press)
and references therein 

\bibitem{SIEG94}
Siegert M and Plischke M 1994 {\it Phys.\ Rev.\ E\/} {\bf 50} 917

\bibitem{LAND00}
Landau D~P and Binder K 2000 {\em Monte Carlo Simulations in Statistical 
Physics\/} (Cambridge: Cambridge Univ.\ Press) 

\bibitem{SHIM01}
Shim Y and Landau D~P 2001 {\it Phys.\ Rev.\ E\/} {\bf 64} 036110 

\bibitem{KANG89}
Kang H~C and Weinberg W~H 1989 {\it J.\ Chem.\ Phys.\/} {\bf 90} 2824; 
Fichthorn K~A and Weinberg W~H {\it ibid\/} {\bf 95} 1090

\bibitem{SHOC92}
Shochet O, Kassner K, Ben-Jacob E, Lipson S~B and M{\"u}ller-Krumbhaar H 
1992 Physica A {\bf 181} 136
and references therein

\bibitem{BURT51}

Burton W~K, Cabrera N and Frank F~C
1951 {\em Phil.\ Trans.\ Roy.\ Soc.\ (London) Ser.\ A\/} {\bf 243}  299

\bibitem{RIKV00}
Rikvold P~A and Kolesik M 2000 {\em J.\ Stat.\ Phys.\/} {\bf 100} 377

\bibitem{RIKV02}
Rikvold P~A and Kolesik M in preparation

\bibitem{KARD86}
Kardar M, Parisi G and Zhang Y-Z 1986 {\em Phys.\ Rev.\ Lett.\/} {\bf 56} 889 

\bibitem{JIAN92}
Jiang Z and Ebner C 1992 {\it Phys.\ Rev.\ B\/} {\bf 45} 6163 

\bibitem{MARR99}
Marro J and Dickman R 
1999 {\em Nonequilibrium Phase Transitions in Lattice Models\/} 
(Cambridge: Cambridge Univ.\ Press) ch~7

\bibitem{GROS91}
Guo H, Grossmann B and Grant M 1990 {\em Phys.\ Rev.\ Lett.\/} {\bf 64} 1262

\bibitem{KOTR91}
Kotrla M and Levi A~C 1991 {\em J.\ Stat.\ Phys.\/} {\bf 64}  579

\bibitem{HONT97}
Hontinfinde F, Krug J and Touzani M 1997 {\em Physica A\/} {\bf 237} 363 

\bibitem{BORT75}
Bortz A~B, Kalos M~H and Lebowitz J~L
1975 {\em J.\ Comput.\ Phys.\/} {\bf 17}  10

\bibitem{MEAK86B}
Meakin P, Jullien R and Botet R
1986 {\em Europhys.\ Lett.\/} {\bf 1} 609

\bibitem{HIRS86}
Hirsch R and Wolf D~E 
1986 {\em J.\ Phys.\ A\/} {\bf 19} L251 

\bibitem{NEER97}
Neergaard J and den~Nijs M
1997 {\em J.\ Phys.\ A\/} {\bf 30} 1935

\end{thebibliography}
\end{document}